# CRISIS ALERT:
## Forecasting Stock Market Crisis Events Using Machine Learning Methods

Xingyi Andrew, Yue Chen, Salintip Supasanya

## 1. Introduction

Historically, the economic recession often came abruptly and disastrously. For instance, during the 2008 financial crisis (the Great Recession), the S&P 500 fell 46.13% from October 2007 to March 2009. Millions of jobs and small businesses were swept away by the depression involuntarily. If we could detect the signals of the crisis earlier, we could have taken preventive measures. Therefore, driven by such motivation, we use advanced machine learning techniques we learned in class, including Random Forest and Extreme Gradient Boosting, to predict any potential market crashes mainly in the US market. Also, we would like to compare the performance of these methods and examine which model is better for forecasting US stock market crashes. Specifically, we are referring to a paper called "Forecasting stock market crisis events using deep and statistical machine learning techniques" as our guidance. We apply our models on the daily financial market data, which tend to be more responsive with higher reporting frequencies. We consider 75 explanatory variables, including general US stock market indexes, S&P 500 sector indexes, as well as market indicators that can be used for the purpose of crisis prediction. Finally, we conclude, with selected classification metrics, that the Extreme Gradient Boosting method performs the best in predicting US stock market crisis events.

## 2. Data Collection and Processing

### 2.1 Data Collection

In order to train the model for stock market crisis prediction, we collect various financial indicators from the Bloomberg Terminal. The dataset includes 5775 observations and covers the period from 01/04/2000 to 02/18/2022, measured on a daily basis.

Our data consists of some major financial crises that have happened in the US, including (Williams, 2022):

- The US dot-com bubble (March 2000)
- Stock market downturn (October 2002)
- Global financial crisis (November 2007)
- Stock market selloff (August 2015)
- Cryptocurrency crash (February 2018)
- Stock market crash (March 2020)

Due to the timeline of listed market crashes, we split our data into two parts: a training dataset, including the data until the end of 2013, and a testing dataset, spanning over the years 2014–2022. The training dataset is used for fitting the model while the testing dataset is used for validating the fitted model.

We examine the following 19 variables to represent different components of the US economy.

| Stock markets | Sector markets | Exchange rates | Additional variables |
| --- | --- | --- | --- |
| Dow Jones Industrial Average | Energy Select Sector | Chinese Yuan to U.S. Dollar | VIX Index |
| S&P 500 | Technology Select Sector | Japan Yen to U.S. Dollar | Gold Price |
| Nasdaq Composite | Financial Select Sector | U.S. Dollars to Euro | Oil Price |
| Hang Seng Index | Consumer Discretionary Select Sector | U.S. Dollars to British Pound | Effective Federal Funds Rate |
|  | Health Care Select Sector |  | Investment-grade Bond Yield |
|  |  |  | US 10-year Bond Yield |

**Table 1.** Examined exploratory variables.

The three US stock market indexes, Dow Jones Industrial Average, S&P 500, and Nasdaq Composite, represent blue-chip, 500 largest market-capitalization, and technology-heavy companies in the US, respectively, which portray the US stock market in the full picture. The Hang Seng Index is a Hong Kong stock market index representing a foreign market that is highly correlated with the US market. The selected five S&P sector markets are important components of the US economy. The four exchange rates are the major players in the foreign exchange market which are influential on the US market. The VIX index measures the volatility of the US stock market, reflecting investors' fear; this index tends to peak when there are crisis events coming. The gold price signals investors' sentiments in which the safe-haven price tends to rise during market uncertainty. The oil price is an important market indicator as high oil prices often negatively impact the US economy. The federal funds rate, investment-grade, and US 10-year bond yields reflect the US economy as bond yields tend to drop during economic contractions.

*2.2. Data processing*

We select our data timeframe to be from January 4, 2000 to February 17, 2022. By following the method proposed in Chatzis et al. (2018), we perform the following data processing procedures:

*Missing Values:* We use backward filling and then forward filling to eliminate the empty values. Since we need to identify the crash the earlier the better, we use the backward first to fill in the gaps with the first value after the missing values. Then, we use the forward filling to take care of those that do not have values available after the missing entries.

*Transformation:* To ensure variables are on the same scale, we calculate daily log-returns (lg_ret) for variables under different S&P 500 sector indexes, general US Stock market indexes, and exchange rates. Also, we include the volatilities of the above log-returns (ln2).

*Dependent variables:* Following the paper's definition ([Chatzis et al., 2018](#)), we create a binary variable that measures whether there is a significant crisis event in the next trading day. We define a "crisis event" for the US market at each trading day if the log-return of the stock index (Dow Jones, S&P 500, or Nasdaq Composite) was below the first percentile of the associated empirical distribution of log-returns. We estimate the empirical distributions of the log-returns using the method from [Bell (2017)](#). We calculate the initial empirical distribution of log-returns based on the first 200 observations, covering the period 1/4/2000–10/9/2000. For each day forward, we recalculate the empirical distribution of log-returns to account for the new observation. If the log-return of that new day was below the first percentile of the new empirical distribution, then we mark it as "crisis". Thus, for the last observation (i.e. 2/17/2022), the empirical distribution of log-returns was based on the period: 1/4/2000–2/16/2022. By using such a method, we derive a two-class dependent variable: crisis events and non-crisis events. Then we shift the binary variable one day ahead to create the one-day prediction horizon. From [Fig.1](#), it is obvious that the dependent variable is highly imbalanced since crises are rare.

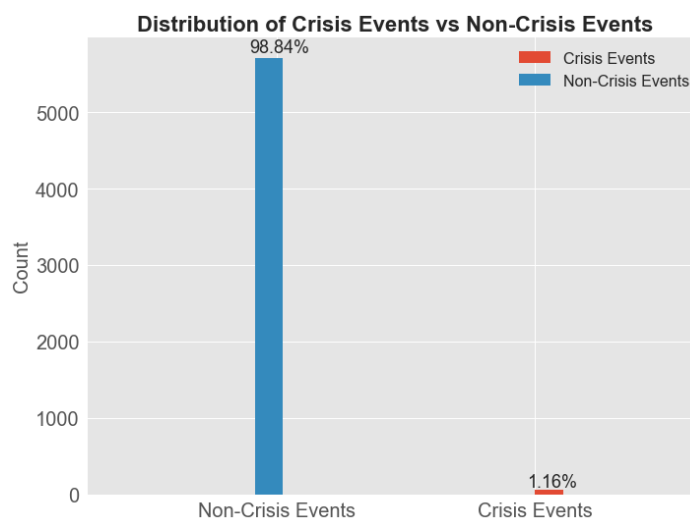

**Fig.1** Distribution of crisis events and non-crisis events (Imbalance Data).

*Lagged variables:* We create 1-day to 5-day lagged variables for fed fund rate, investment-grade bond, US 10-year yield, and exchange rates as we believe these features will have lagged effects on the development of crisis events. We did not include the lagged variables at first, but the model performance was highly inconsistent. Thus, we create additional lagged variables to empower the models' predictabilities.

Our final time frame is from 1/10/2000 to 2/17/2022 after dropping the missing entries caused by lagged variables. We perform a train-test split using the reason discussed above. The train time period is from 1/10/2000 to 12/31/2013 and the test time period is from 1/1/2014 to 2/17/2022.

*2.3. Variable selection*

Since we have 75 explanatory variables, fitting a model to too many variables is prone to the curse of dimensionality. To reduce such effect and improve the model performance, we decide to implement the LASSO regression to perform feature selection, as introduced in the paper ([Chatzis et al., 2018](#)). LASSO is a regression method that penalizes the number of parameters to perform variable selection and regularization to enhance prediction accuracy. We perform the LASSO analysis using the GLMNET package in R which we learned in Machine Learning I. This method efficiently selects the best model configuration from high dimensional data by minimizing the penalized parameter with the cross-validation method, which narrows down to a set of 25 selected features ([Appendix A](#)). The resulting fitted model using selective features will be more regularized for the test data and will reduce overfitting tendencies.

## 3. Model Development

Bagging and boosting are two widely used ensemble learners, thus we perform Random Forests and Extreme Gradient Boosting methods to predict the US market crises. We provide an in-depth explanation of the model development as well as the parameter tuning process for each model.

*3.1 Random Forests (RF)*

Random Forests is a machine learning technique that is widely used to solve regression and classification problems. Since we are investigating a binary dependent variable, we employ the RandomForestClassifier function in the sklearn package in Python. The steps involved in the RF algorithm are as follows: we first create different random subsamples from the sample data with replacement, as known as bootstrapping. Each subsample forms a decision tree and each tree generates a predicted class. The final output is chosen based on a majority vote, called bagging.

To improve the overall performance of our model, we need to find the optimal parameters by means of a tuning algorithm. For this matter, we resort to k-fold cross-validation, which is a technique that employs a limited sample to estimate how the model is expected to perform when used on unseen data. Here, we set the number of folds k equal to 5. We take a range of values for hyperparameters, including the number of trees in the forest, the minimum size of a node before the node is split, the maximum number of features considered for splitting a node, the maximum depth of generated trees, and the minimum leaf node size ([Koehrsen, 2018](#)).

To compare the performance of the models, we set the AUC score, which is the area under the ROC curve, as the scoring objective when doing cross-validation. The AUC score indicates how well a classifier is doing at distinguishing two classes. The higher the AUC, normally should be greater than 0.5, the more accurate the model is at predicting classes. Eventually, we discover a set of parameters that results in a model with the highest AUC score.

Before tuning, the AUC score is 0.9675. After tuning, the AUC score increases to 0.9814, showing that we improve the predictive accuracy. We detect overfitting if the area between the ROC curve of the model predicted on training data and the ROC curve of the model predicted on the testing data is large. As observed in [Appendix B](#), the overfitting problem appears to be eased as that area is smaller for the after-tuning model than for the before-tuning model.

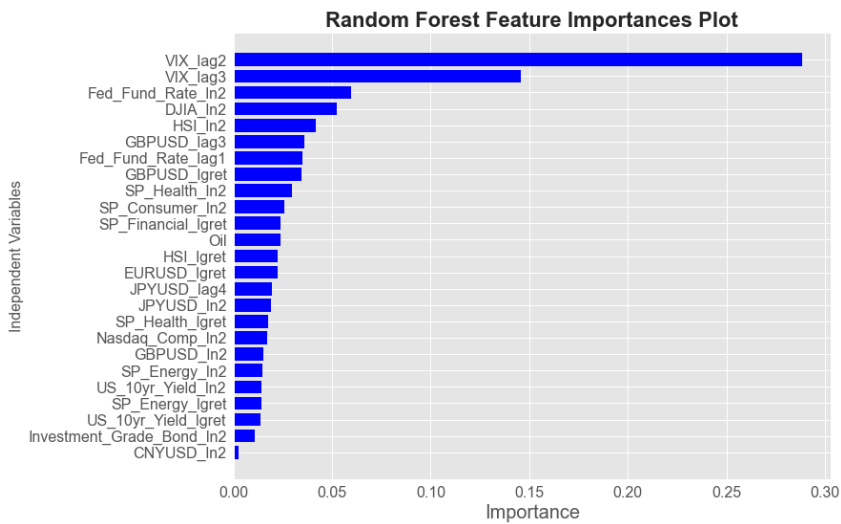

**Fig. 2.** Random forest variable importance plot (Impurity-based)

In Fig. 2, we present impurity-based feature importances for the classification outcome in the RF method. In the RF algorithm, the feature importance is calculated by the average decrease in impurity across all decision trees (Pedregosa et al., 1970). A higher value indicates a feature is more important. This chart indicates that the two-day and three-day lagged values of VIX, the volatility of the federal fund rate as well as the major U.S. indexes achieve higher importance among 25 features, which impose higher predictive power of a crisis event in the US market. It is intriguing to observe that a crisis event is persistent as lagged variables gain greater importance.

*3.2. Extreme Gradient Boosting (XGBoost)*

Extreme Gradient Boosting (XGBoost) is also a powerful machine learning technique that is widely used to solve classification problems. We adopt the XGBoost algorithm, using the xgboost package in Python, to perform predictions of US market crises.

XGBoost uses a boosting method in which shallow trees are built sequentially such that the next tree aims to reduce the error of the previous tree. Eventually, these weak learners will be combined to create a strong learner. In contrast to bagging or RF, in which we construct deep trees to stay unbiased and average trees to reduce variance, the boosting method uses shallow trees with low variance and utilizes sequential learning to decrease bias.

The XGBoost requires more tuning than random forests. Based on the method (Aarshay, 2016), we sequentially perform a 5-fold cross-validation procedure to select a set of entailed hyperparameters, including the maximum depth of generated trees, the minimum leaf node size for performing a split, the minimum weight to create a new node, the size of subsampling, the number of variables in each split, the balance of positive and negative weights and the maximum delta step we allow each leaf output to be. To reduce overfitting tendencies, we tune the gamma hyperparameter, which controls the loss reduction to make a further partition on a leaf node. We also tune alpha and lambda, which control regularization on model weights.

Before tuning, the AUC score is 0.9862. After tuning, the AUC becomes 0.9627. Although the AUC decreases, we observe from Appendix B that the overfitting issue is significantly reduced.

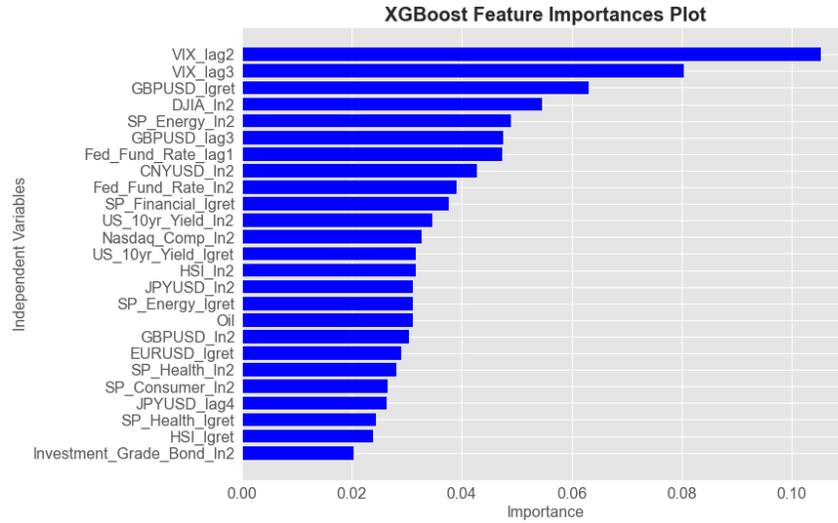

**Fig. 3.** XGBoost variable importance plot.

In Fig. 3, we show feature importances for the outcome in the XGBoost model. This chart indicates that the two-day and three-day lagged values of VIX, the volatility of the federal fund rate, the log-return of GBPUSD exchange rates as well as the major U.S. indexes achieve higher importance among 25 features, showing high predictive power on crisis events.

**4. Experimental Evaluation**

*4.1 Validation Measures*

As Fig. 1 shows that our data is highly imbalanced (98.84%-1.16%), we adopt an asymmetric loss function by adjusting the threshold of predicting whether there is a crisis event. The ultimate goal of this project is to produce an alarming tool that correctly signals the crisis as much as possible. We believe missing a crisis that actually happened (false negative) would be more harmful than predicting a crisis that never existed (false positive).

> True Positive (TP): the number of crises that are correctly identified as crisis
> True Negative (TN): the number of noncrises that are correctly identified as noncrisis
> False Negative (FN): the number of crises that are incorrectly identified as noncrisis
> False Positive (FP): the number of non-crisis that are incorrectly identified as crisis

Therefore, we weigh 20 times higher the cost of false negatives compared to the cost of false positives. We use the following threshold to classify the crisis label.

$$threshold = \frac{FP\ loss}{FN\ loss + FP\ loss} = \frac{1}{21}$$

As a result, any prediction probability that exceeds this threshold will be classified as a crisis event. The confusion matrices presented (Table. 2) are based on this loss function.

| **Random Forest** | Predicted Crisis | Predicted Noncrisis | **XGBoost** | Predicted Crisis | Predicted Noncrisis |
|---|---|---|---|---|---|
| True Crisis | 18 | 1 | True Crisis | 12 | 7 |
| True Noncrisis | 110 | 1993 | True Noncrisis | 4 | 2099 |

**Table.2** Comparison of the confusion matrix of Random Forest and XGBoost after tuning.

In order to evaluate the performance of the two models, we decide to adopt some metrics from Chatzis et al. (2018) to examine models. Due to the imbalanced nature of our data, classification accuracy may be misleading sometimes, we hence look into the following metrics to interpret the contribution of each class to the total accuracy of the fitted models.

$$\text{Sensitivity} = \frac{TP}{TP+TN} \qquad \text{Specificity} = \frac{TP}{TP+TN}$$

Discriminant power (DP): a measure that summarizes sensitivity and specificity. DP values higher than 3 indicate that the algorithm distinguishes well between crisis and non-crisis events.

$$DP = \frac{\sqrt{3}}{\pi}[log\frac{Sensitivity}{1-Sensitivity} + log\frac{Specificity}{1-Specificity}]$$

Balanced accuracy (BA): the average of Sensitivity and Specificity. This term takes into equal consideration the accuracy of the model in the majority (noncrisis) and the minority class (crisis).

$$BA = \frac{1}{2}(Sensitivity + Specificity)$$

Weighted balanced accuracy (WBA): this weights sensitivity more than specificity under the weighting scheme 65%-35%.

$$WBA = 65\%\, Sensitivity + 35\%\, Specificity$$

*4.2 Quantitative Results*

|  | AUC Score | BA | WBA | DP |
|---|---|---|---|---|
| **Random Forest** | 0.981355 | 0.570062 | 0.441231 | 3.190699 |
| **XGBoost** | 0.962660 | 0.873338 | 0.836337 | 3.750094 |

**Table.3** Comparison of the evaluation metrics of RF and XGBoost.

From Table.3, we can observe that only the AUC score of the XGBoost model is lower than that of the RF model. All other metrics for the XGBoost model are higher than those for RF. From the ROC Curves comparing the RF and XGBoost (Fig. 4), we can observe that the XGBoost performs better from false-positive rate (FPR) = 0.1 and all the way to the right, while the RF does better when FPR is smaller than 0.1. From Appendix B, we notice that the RF model carries a greater overfitting tendency than the XGBoost does. Although this contradicts our anticipation because RF is usually robust to overfitting, we believe this issue might occur because of our extremely imbalanced data. In an RF model, a bootstrap sample may contain few or none of the minority class (crisis), which may result in a tree with poor performance for predicting the minority class (Brownlee, 2021). XGBoost might win in this case because it learns its failure in every iteration and improves its ability to predict the class with low participation sequentially.

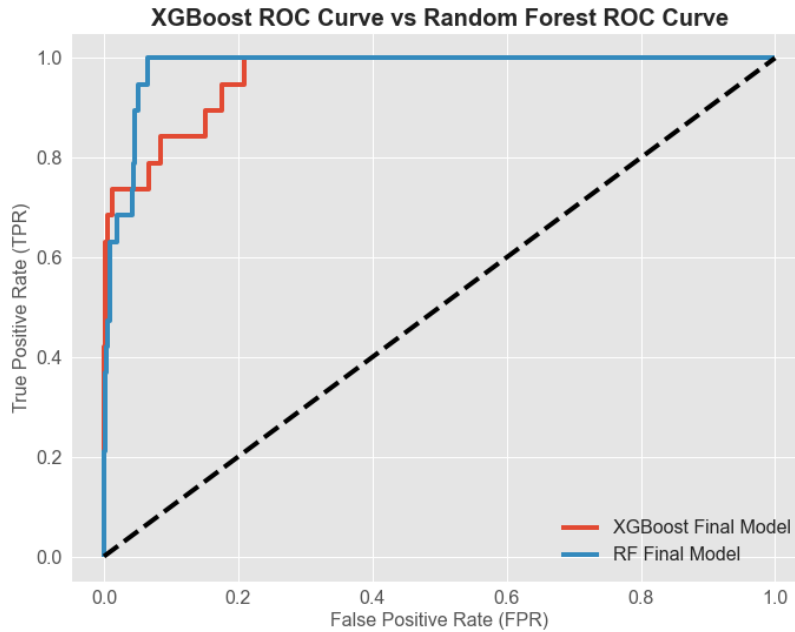

**Fig. 4.** The ROC Curves of RF and XGBoost.

## 5. Conclusion

In summary, we conclude that the XGBoost method has better performance than the RF method. Although the AUC score of the XGBoost model is lower than that of the RF model, all other classification metrics display the superiority of XGBoost in overall prediction accuracy. Moreover, unlike the RF, XGBoost has a negligible overfitting issue, as the ROC curve is consistent from the training set to the testing set. Therefore, we believe that the XGBoost we develop would be a better choice to perform the one-day ahead market crisis prediction. From the feature importance plots (Figs. 2 and 3), we observe that the VIX index that is lagged for two days and three days has the highest importance on both models, suggesting that the VIX index has predictive power on market crashes. This aligns with our expectations since investors' fears usually reflect what the market is facing in the near future.

*Possible improvement:* Since we only include financial data in our model, we might not be able to identify the crisis that is not caused by financial-related issues. We could consider including more non-financial data, such as COVID-related data. Furthermore, adopting Natural Language Processing (NLP) tools to process news to include data from social aspects could be another possible improvement. From the paper (Chatzis et al. 2018), the authors also applied Neural Network and concluded that it has the best performance. This could be another possible approach for us to employ our research specifically focusing on the prediction of the US market.

**Appendix**

Appendix A. K-Fold Cross Validation Test for Feature Selection

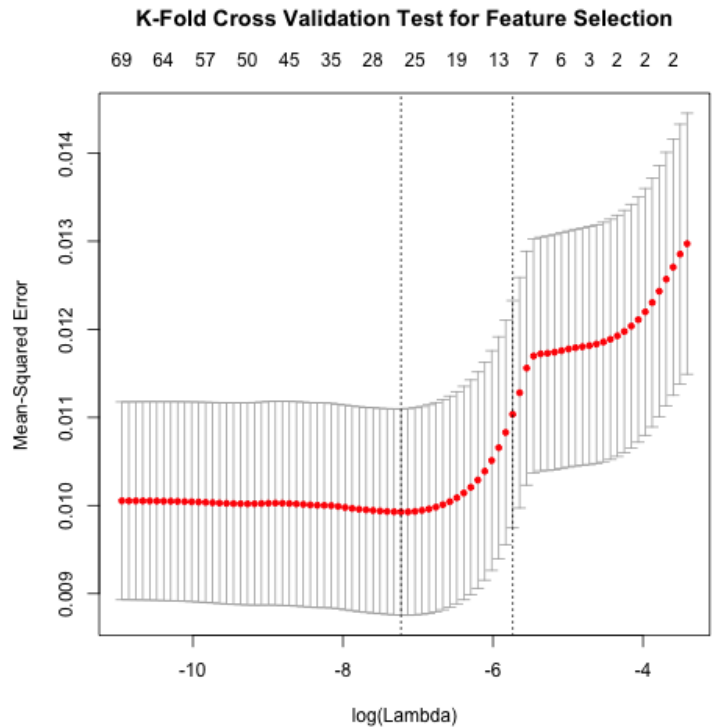

The variable list selected by LASSO using the optimal lambda:

| SP_Energy_lgret | EURUSD_lgret | DJIA_ln2 | US_10yr_Yield_ln2 | VIX_lag2 |
|---|---|---|---|---|
| SP_Financial_lgret | GBPUSD_lgret | Nasdaq_Comp_ln2 | CNYUSD_ln2 | VIX_lag3 |
| SP_Health_lgret | SP_Energy_ln2 | HSI_ln2 | JPYUSD_ln2 | GBPUSD_lag3 |
| HSI_lgret | SP_Consumer_ln2 | Fed_Fund_Rate_ln2 | GBPUSD_ln2 | JPYUSD_lag4 |
| US_10yr_Yield_lgret | SP_Health_ln2 | Investment_Grade_Bond_ln2 | Fed_Fund_Rate_lag1 | Oil |

Appendix B. The RF's ROC Curves for Training Set vs. Testing Set

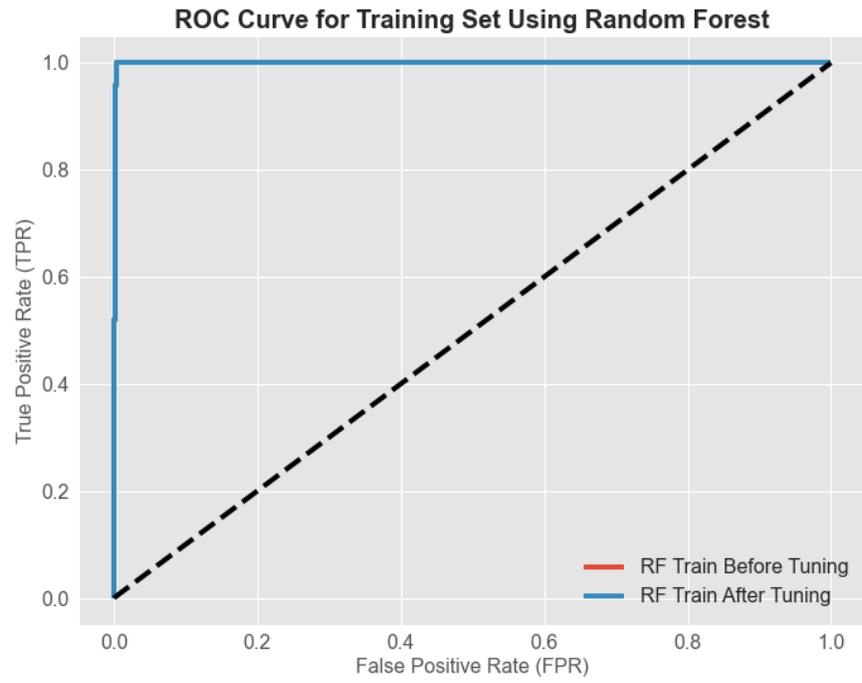

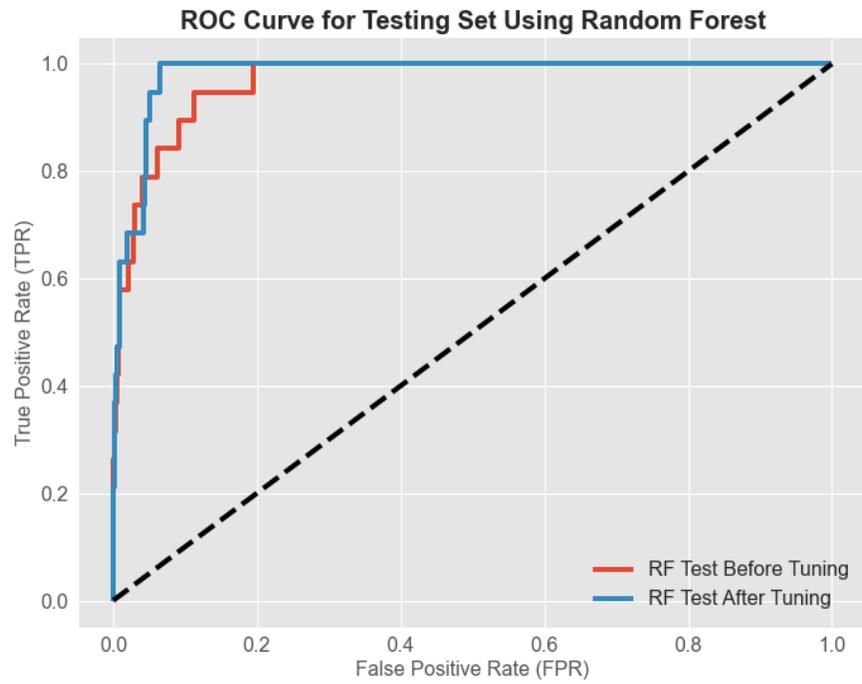

Appendix C. The XGBoost's ROC Curves for Training Set vs. Testing Set

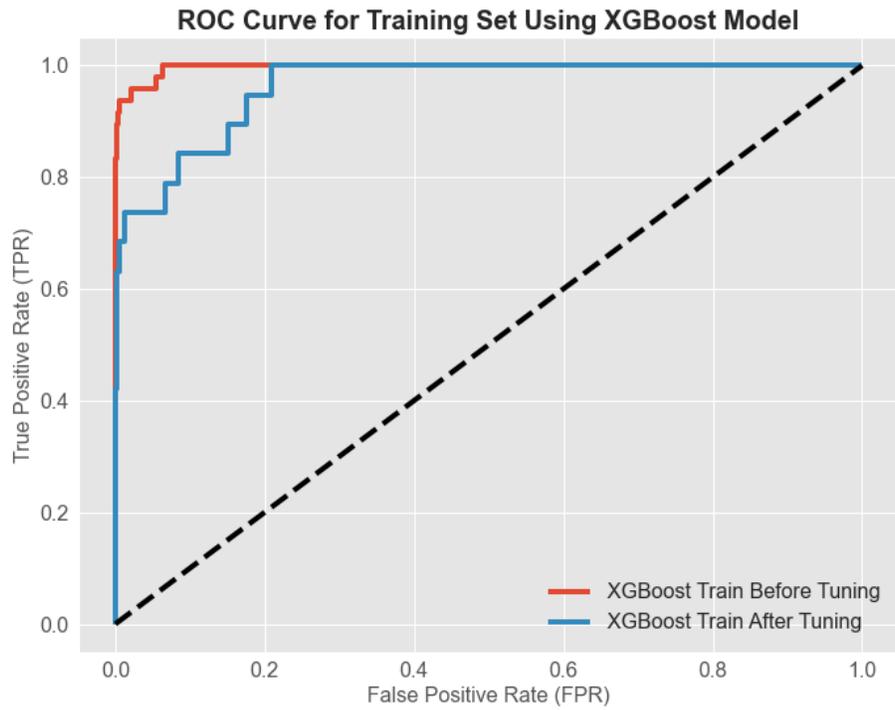

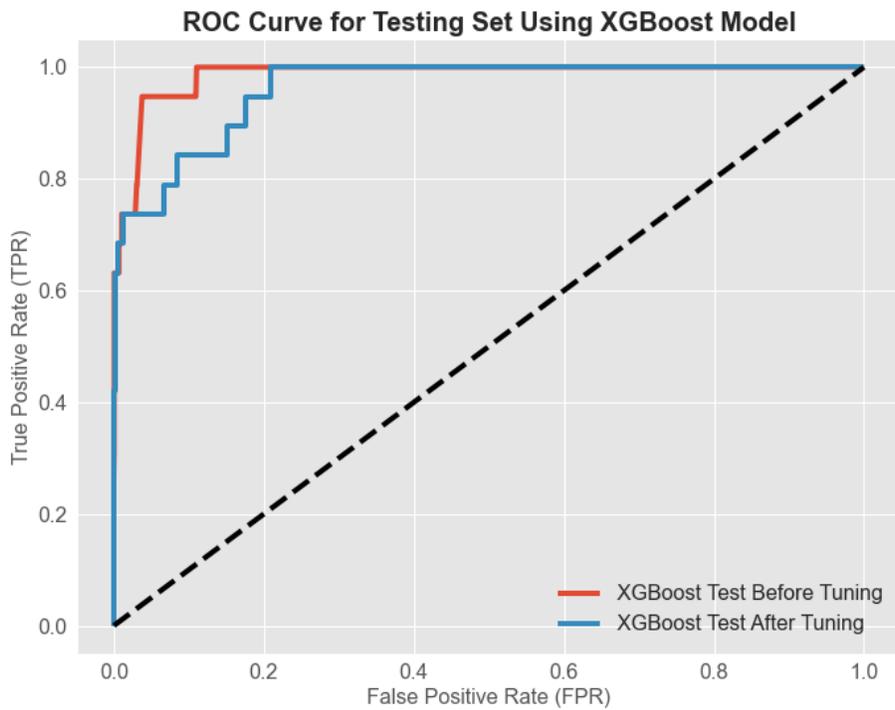

Appendix D. Detailed Variable List

| Variable | Bloomberg-specific Ticker |
|---|---|
| Dow Jones Industrial Average | INDU |
| S&P 500 | SPX |
| Nasdaq Composite | CCMP |
| Hang Seng Index | HSI |
| S&P Energy Select Sector | XLE |
| S&P Technology Select Sector | XLK |
| S&P Financial Select Sector | XLF |
| S&P Consumer Discretionary Select Sector | XLY |
| S&P Health Care Select Sector | XLV |
| Chinese Yuan to U.S. Dollar | CNYUSD |
| Japan Yen to U.S. Dollar | JPYUSD |
| U.S. Dollar to Euro | EURUSD |
| U.S. Dollar to British Pound | GBPUSD |
| VIX Index | VIX |
| Gold Price | XAU Curncy |
| Oil Price | CL1 Commodity |
| Effective Federal Funds Rate | FEDL01 |
| Barclays Investment-grade Bond Yield | LBUSTRUU |
| US 10-year Bond Yield | USGG10YR |

1. Different sector indexes of the S&P 500 market
    a. S&P Energy Select Sector
        i. consists of all companies that play a part in the oil, gas, and consumable fuels business
        ii. reflects the **industrial growth** of an economy

        b. S&P Technology Select Sector
            i. consists of companies that develop or distribute technological items or services, and includes internet companies
            ii. often reflects the most attractive **growth investments** in an economy
        c. S&P Financial Select Sector
            i. includes all companies involved in finance, investing, and the movement or storage of money
            ii. the **primary driver** of a country's economy
        d. S&P Consumer Discretionary Select Sector
            i. are luxury items or services that are not necessary for survival
            ii. reflects **economic growth** phase
        e. S&P Health Care Select Sector
            i. consists of medical supply companies, pharmaceutical companies, and scientific-based operations or services that aim to improve the human body or mind
2. General US stock market indexes
    a. The Dow Jones Industrial Average
        i. a stock market index that tracks 30 large, publicly-owned **blue-chip** companies trading on the New York Stock Exchange (NYSE) and the Nasdaq
        ii. the most-watched stock market indexes in the world
    b. The S&P 500 Index
        i. a stock market index tracking the performance of 500 large companies listed on stock exchanges in the United States
        ii. one of the most commonly followed equity indices
    c. The Nasdaq Composite Index
        i. the market capitalization-weighted index of over 3,000 common equities listed on the Nasdaq stock exchange
        ii. composition is over 50% **technology**, with consumer services, consumer goods, and financials the next most prominent industries
    d. Foreign country:
        i. the main indicator of the overall market performance in Hong Kong
3. Market indicators
    a. VIX index
        i. a real-time index that represents the market's expectations for the relative strength of near-term price changes of the S&P 500 index
    b. Gold price
    c. Oil price
    d. Effective Federal funds rate
    e. Barclays US Investment-grade bond yield
    f. US 10-year bond yield
4. Exchange Rate
    a. CNYUSD: Chinese Yuan to U.S. Dollar
    b. JPYUSD: Japan Yen to U.S. Dollar
    c. EURUSD: U.S. Dollars to Eur
    d. GBPUSD: U.S. Dollars to British Pound